\documentclass[longauth]{aa}
\usepackage{graphicx}
\newcommand{\abel}{Abell~496}

\begin{document}

\title{Very high energy gamma-ray observations of the galaxy clusters Abell~496 and Abell~85 with H.E.S.S.}

\author{F. Aharonian\inst{1,13}
 \and A.G.~Akhperjanian \inst{2}
 \and G.~Anton \inst{16}
 \and U.~Barres de Almeida \inst{8} \thanks{supported by CAPES Foundation, Ministry of Education of Brazil}
 \and A.R.~Bazer-Bachi \inst{3}
 \and Y.~Becherini \inst{12}
 \and B.~Behera \inst{14}
 \and K.~Bernl\"ohr \inst{1,5}
 \and C.~Boisson \inst{6}
 \and A.~Bochow \inst{1}
% ...
}

\newpage

\institute{
Max-Planck-Institut f\"ur Kernphysik, P.O. Box 103980, D 69029
Heidelberg, Germany
\and
 Yerevan Physics Institute, 2 Alikhanian Brothers St., 375036 Yerevan,
Armenia
\and
Centre d'Etude Spatiale des Rayonnements, CNRS/UPS, 9 av. du Colonel Roche, BP
4346, F-31029 Toulouse Cedex 4, France
\and
Universit\"at Hamburg, Institut f\"ur Experimentalphysik, Luruper Chaussee
149, D 22761 Hamburg, Germany
\and
Institut f\"ur Physik, Humboldt-Universit\"at zu Berlin, Newtonstr. 15,
D 12489 Berlin, Germany
\and
LUTH, Observatoire de Paris, CNRS, Universit\'e Paris Diderot, 5 Place Jules Janssen, 92190 Meudon, 
France
Obserwatorium Astronomiczne, Uniwersytet Ja
\and
IRFU/DSM/CEA, CE Saclay, F-91191
Gif-sur-Yvette, Cedex, France
\and
University of Durham, Department of Physics, South Road, Durham DH1 3LE,
U.K.
\and
Unit for Space Physics, North-West University, Potchefstroom 2520,
    South Africa
\and
Laboratoire Leprince-Ringuet, Ecole Polytechnique, CNRS/IN2P3,
 F-91128 Palaiseau, France
\and 
Laboratoire d'Annecy-le-Vieux de Physique des Particules, CNRS/IN2P3,
9 Chemin de Bellevue - BP 110 F-74941 Annecy-le-Vieux Cedex, France
\and
Astroparticule et Cosmologie (APC), CNRS, Universite Paris 7 Denis Diderot,
10, rue Alice Domon et Leonie Duquet, F-75205 Paris Cedex 13, France
\thanks{UMR 7164 (CNRS, Universit\'e Paris VII, CEA, Observatoire de Paris)}
\and
Dublin Institute for Advanced Studies, 5 Merrion Square, Dublin 2,
Ireland
\and
Landessternwarte, Universit\"at Heidelberg, K\"onigstuhl, D 69117 Heidelberg, Germany
\and
Laboratoire de Physique Th\'eorique et Astroparticules, CNRS/IN2P3,
Universit\'e Montpellier II, CC 70, Place Eug\`ene Bataillon, F-34095
Montpellier Cedex 5, France
\and
Universit\"at Erlangen-N\"urnberg, Physikalisches Institut, Erwin-Rommel-Str. 1,
D 91058 Erlangen, Germany
% ...
}

\offprints{\email{Wilfried.Domainko@mpi-hd.mpg.de, Dalibor.Nedbal@mpi-hd.mpg.de}}

\authorrunning{H.E.S.S. Collaboration}
\titlerunning{H.E.S.S. observations of Abell~496 and Abell~85}
\date{Received / Accepted}
 
\abstract
{}
{The nearby galaxy clusters Abell~496 and Abell~85 are studied in the very high energy (VHE, E$>$ 100 GeV) band to investigate VHE cosmic rays (CRs) in this class of objects which are the largest gravitationally bound systems in the Universe.}
{H.E.S.S., an array of four Imaging Atmospheric Cherenkov Telescopes (IACT), is used to observe the targets in the range of VHE gamma rays.}
{No significant gamma-ray signal is found at the respective position of the two clusters with several different source size assumptions for each target. In particular, emission regions corresponding to the high density core, to the extension of the entire X-ray emission in these clusters, and to the very extended region where the accretion shock is expected, are investigated. Upper limits are derived for the gamma-ray flux at energies E$>570$ GeV for Abell~496 and E$>460$ GeV for Abell~85.}
{From the non-detection in VHE gamma rays, upper limits on the total energy of hadronic CRs in the clusters are calculated. If the cosmic-ray energy density follows the large scale gas density profile, the limit on the fraction of energy in these non-thermal particles with respect to the total thermal energy of the intra-cluster medium (ICM) is 51\% for Abell~496 and only 8\% for Abell~85 due to its larger mass and higher gas density. These upper limits are compared with theoretical estimates. They predict about $\sim$10\% of the thermal energy of the ICM in non-thermal particles. The observations presented here can constrain these predictions especially for the case of the Abell~85 cluster.}


\maketitle


\section{Introduction}

Clusters of galaxies, the most extended gravitationally bound systems in the Universe, are expected to contain a significant population of non-thermal particles. 
The mechanisms for producing cosmic rays (CR) in galaxy clusters can be divided into external processes and internal processes. 
In external processes, the particle acceleration is driven by the assembly of the cluster. An efficient production of high energy particles is expected especially at the strong accretion shock at the outskirt of the cluster, where cold in-falling material plunges into the already existing hot intra-cluster medium (ICM). Therefore, large-scale shock waves caused by the cosmological structure formation may populate these objects with a non-thermal component of particles (see e.g. Colafrancesco \& Blasi \cite{colafrancesco00}, Loeb \& Waxman \cite{loeb00}, Ryu et al. \cite{ryu03}, Miniati \cite{miniati03}).
Particles can also be accelerated by turbulence in the intra-cluster medium(ICM) generated by major sub-cluster merger events (e.g. Brunetti et al. \cite{brunetti04}).
In contrast in the internal mechanisms, the CRs are accelerated by cluster galaxies and injected into the whole cluster volume afterwards.
Internal sources of CRs in clusters can be supernova-driven galactic winds (V\"olk et al. \cite{voelk96}) or AGNs (e.g. En\ss lin et al. \cite{ensslin97}, Aharonian \cite{aharonian02}, Hinton et al. \cite{hinton07}).

Accelerated, highly energetic CR particles can subsequently produce gamma rays through two main mechanisms in galaxy clusters. VHE gamma rays can be generated by hadronic processes, through inelastic collisions between high-energy protons and target protons provided by the thermal hot ICM and subsequent $\pi^0$-decay (Dennison \cite{dennison80}). In the leptonic scenario, very high energy (VHE, E~$>$~100 GeV) electrons up-scatter cosmic microwave background (CMB) photons to gamma-ray energies via the inverse Compton process (e.g. Atoyan \& V\"olk \cite{atoyan00}, Gabici \& Blasi \cite{gabici03}, \cite{gabici04}). VHE electrons which up-scatter low energy photons to the gamma-ray range can also be of secondary origin either due to photon - photon interactions (Timokhin et al. \cite{timokhin04}) or generated by ultra-high energy protons (E $>$ 10$^{18}$ eV) interacting with CMB photons in the Bethe-Heitler process (Inoue et al. \cite{inoue05}).
 
Galaxy clusters are huge structures with a linear scale of several Mpc. This length scale together with typical cluster magnetic fields in the $\mu$G range (Carilli \& Taylor \cite{carilli02}, Govoni \& Feretti \cite{feretti04}) leads to a large diffusion timescale. This diffusion time is longer than the Hubble time for CR protons with energies less than $\sim$10$^{15}$ eV. Therefore no losses of such particles will occur in these systems (V\"olk et al. \cite{voelk96}, Berezinsky et al. \cite{berezinsky97}). Gamma rays produced through the hadronic scenario can thus be used to probe particle acceleration in clusters of galaxies over the entire formation history.

In contrast to hadronic CRs, electrons which are accelerated to VHE energies in clusters will suffer from substantial energy losses due to synchrotron and inverse Compton radiation. Therefore, an electron with an energy of $\sim$1 TeV has only a limited lifetime of $\sim$10$^6$ years in a typical cluster magnetic field of 1 $\mu$G (e.g. Atoyan \& V\"olk \cite{atoyan00}). Consequently VHE electrons do not accumulate in clusters and it is expected that the electron spectrum will cut off at an energy that is below the VHE regime outside the acceleration sites. Hence leptonic VHE gamma-ray production is likely less relevant in galaxy clusters (Atoyan \& V\"olk \cite{atoyan00}). 

Observationally, a significant population of non-thermal electrons is found in several clusters that can be detected in radio waves (Giovannini \& Feretti \cite{giovannini00}, Feretti et al. \cite{feretti04}, Bagchi et al. \cite{bagchi06}) and possibly hard X-rays (Rephaeli \& Gruber \cite{rephaeli02}, Fusco-Femiano et al \cite{fusco04}). Given 
the evidence/hints of accelerated particles in galaxy clusters, these objects are potential sources
for observation of gamma-rays (see Blasi et al. \cite{blasi07} for a recent review). However, despite these indications, no cluster has been established as a gamma-ray emitter so far (Reimer et al. \cite{reimer03}). In the VHE gamma-ray range Perkins et al. (\cite{perkins06}) have reported upper limits for two nearby massive clusters (Perseus and Abell~2029) with the \textit{Whipple} telescope. For the Perseus cluster, these results imply that the cosmic-ray proton energy density is less than 8\% of the thermal energy density. 

Throughout the paper a standard cosmology with $\Omega_{\Lambda}$~=~0.7, $\Omega_{m}$~=~0.3 and $H_0$~=~70~km~s$^{-1}$~Mpc$^{-1}$ is used. 

In this paper, the observations of the galaxy clusters Abell~496 and Abell~85 with the H.E.S.S. array of telescopes in the VHE gamma-ray regime are presented.

\section{Target overview}

The target clusters were selected in terms of optimal detectability, position and distance for an observation with H.E.S.S. Promising targets of this kind should be located on the southern hemisphere and at a redshift not larger than z $\sim$ 0.06, since more distant objects suffer substantial absorption from extragalactic background light (EBL). Furthermore, there should be no blazar at the location of the cluster which could superpose potential VHE emission of the galaxy cluster. Two of the clusters that were chosen for an observation with H.E.S.S. are presented in this paper. These clusters were selected according to different criteria as described below. The properties of both selected clusters can be seen in Tab. \ref{clusters}\footnote{Quantities found by other authors using a different cosmology are scaled to the value of $H_0$ adopted in this work throughout the paper}.

As a first target, a compact galaxy cluster was selected. 
The sensitivity of imaging atmospheric Cherenkov telescopes decreases approximately with the square root of the solid angle of the gamma-ray emission region, and therefore linearly with the source extension. For this reason the detectability of a source is proportional to its gamma-ray luminosity $F_{\gamma}$ divided by its size $R_{\gamma}$. If it is assumed that the X-ray size and the X-ray brightness of a cluster together are a measure of its gamma-ray flux, then $F_X/R_X$ can be used as a figure of merit. It has to be noted that this selection procedure prefers galaxy clusters that host a so-called \textit{cooling core} at their center. In a cooling core cluster, the central gas density is large enough that the radiative cooling time due to thermal X-ray emission is shorter than the Hubble time (see Peterson \& Fabian \cite{peterson06} for a review). This large density of target material is favorable for hadronic production of gamma rays. However, since only a small fraction of the total gas mass is contained in the cooling core, this will increase the total gamma-ray luminosity by only a modest amount. These selection criteria were applied to the galaxy clusters of the REFLEX survey (B\"ohringer et al. \cite{boehringer04}), which covers two thirds of the southern sky and contains 447 X-ray bright clusters. Based on this selection procedure, Abell~496 was found to be a promising candidate for H.E.S.S. observations.  

Abell~496 is a nearby (z = 0.033) relaxed cluster that features such a \textit{cooling core} at its center. 
It also shows \textit{cold front} substructures (Dupke \& White \cite{dupke03}). Cold fronts are sharp edges in the X-ray surface brightness that are inconsistent with the appearance of shock fronts. In the case of Abell~496, the cold fronts are likely to be caused by the oscillation of the central galaxy around an equilibrium position, since the chemical abundance of the ICM is flat across these structures (Dupke \& White \cite{dupke03}). The absence of a jump in the abundance suggests that material at both sides of the front is of the same origin and hence the cold front is not caused by a sub-cluster merger.

As a second target, a massive cluster was observed with a quite deep exposure. For selecting this target a different procedure was adopted than for Abell~496. Clusters were evaluated according to their accretion power, which scales with M$^{5/3}$ (M is the total mass of the cluster, see Gabici \& Blasi \cite{gabici03}, \cite{gabici04}). Then the estimated accretion power was converted into an `accretion flux' by degrading it by the inverse square of the target distance. Finally the source extension was also taken into account. Since, as described above, the sensitivity of imaging atmospheric Cherenkov telescopes decreases approximately linearly with the source extension, the `accretion flux' was further scaled with the inverse of the cluster radius. The aforementioned criteria were applied to the X-ray  selected galaxy cluster catalog of Reiprich \& B\"ohringer (\cite{reiprich02}). Following this selection procedure the cluster Abell~85 was found to be a promising target.

Abell~85 is a galaxy cluster with a complex morphology at a redshift of 0.055. In X-rays this object shows two sub-clusters merging with the main cluster (Kempner et al \cite{kempner02}, Durret et al. \cite{durret05}). Additionally, Abell~85 features a cooling core at its center which is quite uncommon for merging clusters. Presumably the merging sub-clusters have not yet reached the central region of the cluster and have therefore not disrupted the existing cooling core (Kempner et al. \cite{kempner02}). 

In the adopted cosmological model, the redshift of Abell~496 (Abell~85) corresponds to a distance of 134 Mpc (220 Mpc) and 1$^{\circ}$ relates to 2.35 Mpc (3.86 Mpc) at the target.

\begin{table*}
\begin{center}
\caption{Properties of the two galaxy clusters. L$_\mathrm{X, 0.1-2.4}$ is the luminosity in the 0.1 - 2.4 keV band and R$_{vir}$ is the virial radius. Refs.: a) Reiprich \& B\"ohringer (\cite{reiprich02}), b) Markevitch et al. (\cite{markevitch99}), c) Durret et al. (\cite{durret05}).}\label{clusters}
\begin{tabular}{l|c|c|c|c|c|c|l}
\hline
\hline
cluster & M$_\mathrm{tot}$ [M$_\mathrm{sun}$] & L$_\mathrm{X, 0.1-2.4}$ [erg s$^{-1}$] & T [keV] & R$_\mathrm{vir}$ [Mpc] & redshift & Distance [Mpc] & Refs \\ 
\hline
Abell~496 & 3.1$\times 10^{14}$ & 1.9$\times 10^{44}$ & 4.7 & 1.4 & 0.033 & 134 & a,b \\
Abell~85 & 7.6$\times 10^{14}$ & 4.8$\times 10^{44}$ & 7 & 1.9 & 0.055 & 220 & a,c \\
\hline
\end{tabular}

\end{center}

\end{table*}


\section{Observations and data analysis}
The observations were performed with the H.E.S.S. telescope array, consisting of four IACTs located in Namibia (23$^\circ$16'18'' S 16$^\circ$30'00'' E). The system is described by e.g. \cite{hofmann03}. Data were taken for Abell~496 from October to December 2005 and in October 2006. Abell 85 was observed in October and November 2006 and in August 2007. The observations were performed in \textit{wobble mode} (Aharonian et al. \cite{Aharonian06}) with the target being typically at 0.7$^\circ$ offset from the center of the field-of-view. This allows a simultaneous background estimation in the same field-of-view. In total, 14.6 (32.5) hours of live time on Abell 496 (Abell 85) meet the standard data quality selection criteria and are used for the analysis. The average zenith angle of the observations of Abell 496 (Abell 85) was 28$^\circ$ (18$^{\circ}$) resulting in a post-analysis energy threshold of 570 GeV (460 GeV) for {\em hard cuts} (see Aharonian et al. \cite{Aharonian06}).

The data analysis after calibration consists of a reconstruction of properties of impinging gamma rays through the Hillas-parametrization of the observed Cherenkov images (Hillas \cite{hillas96}), a background subtraction, and an estimation of the flux (Aharonian et al. \cite{Aharonian06}). The direction of the primary gamma photon is estimated using geometrical analysis of multiple camera images according to the algorithm 1 as specified by Hofmann et al. (\cite{hofmann99}). The energy of the primary gamma is estimated by comparing the image intensity with the values predicted by Monte Carlo simulations for the reconstructed gamma-ray direction and impact point (see Aharonian et al. (\cite{Aharonian06}) for details). A detailed comparison of the intensity of Cherenkov images generated by fast muons going through the telescope with simple theoretical calculation provides an absolute calibration of the optical efficiency of each telescope (Aharonian et al. \cite{Aharonian06}).

A system of cuts on parameters of each shower is used to reduce the prevailing hadronic background. {\em Hard cuts} from Aharonian et al. (\cite{Aharonian06}) are used because they were optimized for sources of low flux and for hard spectra. 

The remaining background is estimated using the \textit{reflected regions} background estimation method (\cite{berge07}). The method uses a number of control regions in the same field-of-view (FOV) as the observed target for estimating the background flux. In case of a very extended emitting region (radius $> 0.7$\degr) the background is estimated using the \textit{ON/OFF method} (Weekes et al. \cite{weekes89}). The background is then estimated from off-source runs. To assure an accurate background subtraction, these runs were performed under the same conditions as the on-source runs.

All upper limits are derived using the approach of Feldman and Cousins (Feldman \& Cousins \cite{feldman98}) at a 99.9\% confidence level. A power law gamma-ray spectrum with spectral index $\Gamma=2.1$ is assumed. This value is based upon the expectations of V\"olk et al. (\cite{voelk96}). In order to check the dependence of the results on the assumed index, the upper limits for $\Gamma=2.3$ are also given. The difference is less than 10\% in all cases.

\section{Results}
\label{section_results}
\subsection*{Abell 496}
Since the size of the VHE gamma-ray emitting region is unknown, four different analyses are used, integrating the signal over several sizes of on-source regions at the position of \abel. The integration radii $\theta$ are chosen according to characteristic length scales of the density profile of the ICM, which acts as the target material for hadronic gamma-ray production. The density $\rho$ of the ICM  generally follows a $\beta$ model \cite{cavaliere76}:

\begin{equation}
\rho = \rho_{0}\left[1+ \left( \frac{r}{r_c} \right)^2 \right]^{-3 \beta/2}
\end{equation} 

In the $\beta$ model $\rho_0$ is the central gas density, r$_c$ is the core radius and the parameter $\beta$ describes the slope of the density profile.

The \textit{core analysis} is used to search for a signal coming from the X-ray core region of the cluster. The core size of the $\beta$ model 
was measured by Markevitch et al. (\cite{markevitch99}) as 178 kpc, corresponding to 0.08\degr. This size is comparable to the size of the point-spread-function of H.E.S.S. The signal is therefore integrated over a larger area with radius $\theta_1=0.1$\degr, optimized for a search of point-like sources (Aharonian et al. \cite{Aharonian06}).

\begin{figure}[htbp] 
  \resizebox{\hsize}{!}{\includegraphics{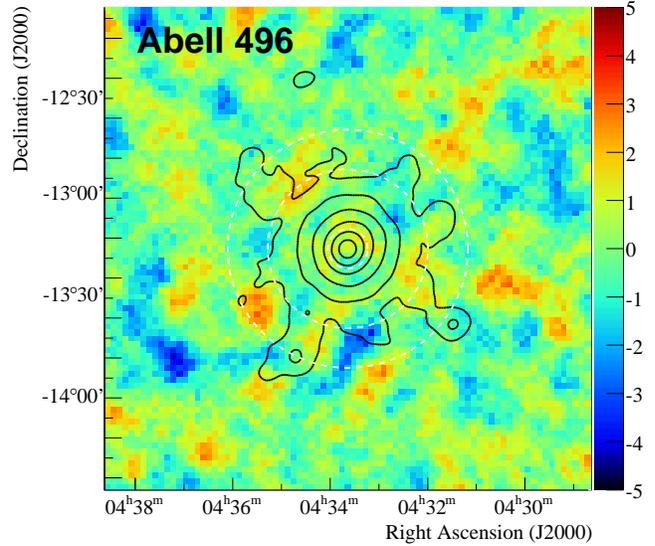}}
\caption{Correlated significance map of the region around Abell~496 obtained using the {\em core analysis}. For each point in the map, the significance is evaluated by counting gamma-ray candidates within a radius of 0.1$^{\circ}$. White dashed circles depict the on-source regions used for the {\em core analysis}, the {\em 1Mpc analysis} and the {\em extended analysis} (corresponding radii $\theta_1=0.1^{\circ}$, $\theta_4=0.4^{\circ}$, $\theta_2=0.6^{\circ}$). The {\em very extended analysis} is not shown since the on-source region is larger than the map. The distribution of significances in the FOV is consistent with background fluctuations. Also shown are black contours from hard-band ROSAT PSPC observations (Durret et al. 2000), smoothed by the H.E.S.S. angular resolution.}\label{fig_sig_dist}
\end{figure}

\begin{figure}[htbp] 
  \resizebox{\hsize}{!}{\includegraphics{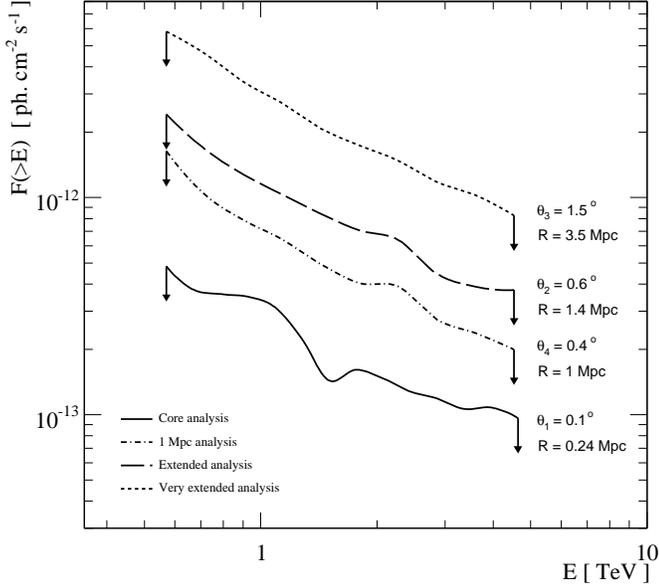}}
  \caption{The H.E.S.S. 99.9\% upper limits on the integral flux from Abell~496 as a function of the energy above which the flux is integrated, assuming four different sizes of the emission region. All curves are obtained by comparing the gamma-ray excess above a given reconstructed energy to the number of events expected for a spectral index of $\Gamma=2.1$. The result for the {\em core analysis} exhibits fluctuations caused by a lower number of excess events. Note that the upper limit for the {\em very extended analysis} was produced with a reduced dataset.}\label{fig_ul}
\end{figure}

No significant emission is found from the central region above the hard cuts threshold energy of 570 GeV. In order to search for other point sources off the center of \abel\, a map was created with significances of point-like TeV gamma-ray signals on a grid around the center of the cluster (Fig. \ref{fig_sig_dist}). The significance
distribution over the FOV is consistent with background fluctuations. The upper limit on the integral flux from the core region is derived to be $F(>570\ \mathrm{GeV}) < 4.8 \times 10^{-13}$ ph. s$^{-1}$ cm$^{-2}$ assuming $\Gamma=2.1$ and $F(>570\ \mathrm{GeV}) < 5.2 \times 10^{-13}$ ph. s$^{-1}$ cm$^{-2}$ for $\Gamma=2.3$, corresponding to 0.9 and 1.0\% of the integral Crab Nebula flux above the same energy as measured by Aharonian et al. (\cite{Aharonian06}). The integral upper limits as a function of the energy above which the flux is shown in Fig. \ref{fig_ul}.

The \textit{extended analysis} is performed on the whole region of the X-ray emission from the cluster, using $\theta_2=0.6$\degr\ as measured by Reiprich \& B\"{o}hringer (\cite{reiprich02}). No signal is found above the energy of 570 GeV. Assuming a photon index of $\Gamma=2.1$, an upper limit on the integral flux is determined $F(>570\ \mathrm{GeV}) < 2.4\times 10^{-12} $ph. s$^{-1}$ cm$^{-2}$, which corresponds to 4.6\% of the integral Crab flux. Figure \ref{fig_ul} compares this result with the {\em core analysis}. Assuming $\Gamma=2.3$, the upper limit is $F(> 570\ \mathrm{GeV}) < 2.6\times 10^{-12} $ph. s$^{-1}$ cm$^{-2}$ (5.0\% Crab flux).

The \textit{very extended analysis} is aimed at investigating possible emission from the accretion shocks. A very large on-source region of radius $\theta_3=1.5$\degr\ is used, corresponding to 3.5 Mpc. The data set is reduced to 9.8 live hours, because not every on-source run had a corresponding good-quality off-source run that could be used for the background estimation. Again, no significant signal is found. An upper limit is determined to be $F(>570\ \mathrm{GeV}) < 5.8\times 10^{-12} $ph. s$^{-1}$ cm$^{-2}$ (10.9\% Crab flux) for $\Gamma=2.1$ and $F(>570\ \mathrm{GeV}) < 6.2\times 10^{-12} $ph. s$^{-1}$ cm$^{-2}$ (11.7\% Crab flux) for $\Gamma=2.3$.

An additional analysis was performed, using the integrating radius $\theta_4=0.4$\degr\ that corresponds to a radius of 1 Mpc from the center of Abell~496. The radius of 1 Mpc is not physically motivated, but other physical quantities that are used in the discussion are well measured within this radius. Hence these results are used for modeling in section \ref{section_discussion}. For the same reason, the upper limits are in this case calculated above 1~TeV. The resulting upper limits are $F(> 1\mathrm{TeV}) < 7.2\times 10^{-13} $ph. s$^{-1}$ cm$^{-2}$ for $\Gamma=2.1$ (3.2\% Crab flux) and $F(> 1\mathrm{TeV}) < 7.5\times 10^{-13} $ph. s$^{-1}$ cm$^{-2}$ (3.3\% Crab flux)for $\Gamma=2.3$. 

All results are summarized in table \ref{table:results}.

\begin{table*}[htbp] 
\begin{center}
\caption{Summary of the results of Abell~496 and Abell~85 observations.}
\label{table:results}
\begin{tabular}{l c c c c c c}
{\bf Abell~496} & & & \\
\hline
\hline
Analysis & Radius       & Radius & $E_{th}$ &Assumed  &  $F_{\rm ul}(>E_{th})$                   & $F_{\rm ul}(>E_{th})$\\
         &[$^{\circ}$]  & [Mpc]  &[TeV]     & $\Gamma$ &  $[10^{-12}~{\rm ph.}~{\rm cm}^{-2}~{\rm s}^{-1}]$ & [\% Crab flux]\\
\hline
{\em Core}     & 0.1 & 0.2 & 0.57 & 2.1 & 0.48 & 0.9 \\
               &     &     &      & 2.3 & 0.52 & 1.0 \\
\hline
{\em 1 Mpc}    & 0.4 & 1.0 & 1.0  & 2.1 & 0.72 & 3.2 \\
               &     &     &      & 2.3 & 0.75 & 3.3 \\
\hline
{\em Extended} & 0.6 & 1.4 & 0.57 & 2.1 & 2.4  & 4.6 \\
               &     &     &      & 2.3 & 2.6  & 5.0 \\
\hline
{\em Very 
extended}      & 1.5 & 3.5 & 0.57 & 2.1 & 5.8  & 10.9 \\
               &     &     &      & 2.3 & 6.2  & 11.7 \\
\hline
\vphantom{\Large{A}}&&&\\
{\bf Abell~85} & & & \\
\hline
\hline
Analysis & Radius       & Radius & $E_{th}$ &Assumed  &  $F_{\rm ul}(>E_{th})$                   & $F_{\rm ul}(>E_{th})$\\
         &[$^{\circ}$]  & [Mpc]  &[TeV]     & $\Gamma$ &  $[10^{-12}~{\rm ph.}~{\rm cm}^{-2}~{\rm s}^{-1}]$ & [\% Crab flux]\\
\hline
{\em Core}     & 0.10 & 0.4 & 0.46 & 2.1 & 0.39 & 0.5 \\
               &      &     &      & 2.3 & 0.41 & 0.6 \\
\hline
{\em 95\% 
X-ray 
containment}   & 0.13 & 0.5 & 0.46 & 2.1 & 0.34 & 0.5 \\
               &      &     &      & 2.3 & 0.36 & 0.5  \\
\hline
{\em 1 Mpc}    & 0.26 & 1.0 & 1.0  & 2.1 & 3.2  & 1.4 \\
               &      &     & 1.0  & 2.3 & 3.3  & 1.4 \\
\hline
{\em Extended} & 0.49 & 1.9 & 0.46 & 2.1 & 1.5  & 2.0 \\
               &      &     &      & 2.3 & 1.6  & 2.2 \\
\hline
{\em Very 
extended}      & 0.91 & 3.5 & 0.46 & 2.1 & 9.9  & 13.6 \\
               &      &     &      & 2.3 & 11.0 & 15.1 \\
\hline
\end{tabular}
\end{center}

\end{table*}

\subsection*{Abell~85}
Similarly to the case of Abell~496, several integration radii $\theta$ were probed to search for a signal from Abell~85. Abell~85 could be observed at higher elevations which led to a lower energy threshold for this target.

\begin{figure}[htbp] 
  \resizebox{\hsize}{!}{\includegraphics{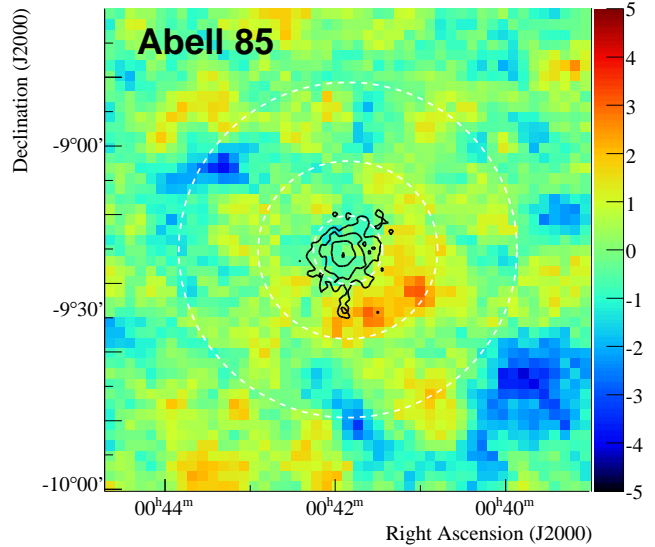}}
\caption{Correlated significance map of the region around Abell~85 obtained using the {\em core analysis} in the same way as in Fig. \ref{fig_sig_dist}.  White dashed circles depict the on-source regions used for the {\em core analysis}, {\em 1 Mpc analysis} and {\em extended analysis} (corresponding radii $\theta_1=0.1^{\circ}$, $\theta_4=0.26^{\circ}$, $\theta_2=0.49^{\circ}$). The distribution of significances in the FOV is consistent with background fluctuations. Also shown are black contours from hard-band ROSAT PSPC observations (Pislar et al. \cite{pislar97}).}\label{fig:a85_sigdist}
\end{figure}

The size of the core region is $226$ kpc (Mohr et al. \cite{mohr99}), corresponding to $\sim 3$'. The {\em core analysis} thus again uses $\theta_1 = 0.1^{\circ}$, the radius optimized for a search for point-like sources. No significant signal is found and an upper limit on the integral flux is derived $F(> 460\ \mathrm{GeV}) < 3.9 \times 10^{-13} $ph. s$^{-1}$ cm$^{-2}$ (0.5\% Crab flux) for $\Gamma=2.1$ and $F(> 460\ \mathrm{GeV}) < 4.1 \times 10^{-13} $ph. s$^{-1}$ cm$^{-2}$ (0.6\% Crab flux) for $\Gamma=2.3$. A significance map of the region around the center of the cluster is shown (Fig. \ref{fig:a85_sigdist}). Also in this case the significance distribution is consistent with background fluctuations.

The {\em extended analysis} probes the region of X-ray overdensity, as measured by Reiprich \& B\"{o}hringer (\cite{reiprich02}) to be $\theta_2 = 0.49^{\circ}$. The integral upper limit in this case is $F(> 460\ \mathrm{GeV}) < 1.5 \times 10^{-12} $ph. s$^{-1}$ cm$^{-2}$ (2.0\% Crab flux) for $\Gamma=2.1$ and $F(> 460\ \mathrm{GeV}) < 1.6 \times 10^{-12} $ph. s$^{-1}$ cm$^{-2}$ (2.2\% Crab flux) for $\Gamma=2.3$.

The \textit{very extended analysis} uses an on-source region of radius $\theta_3=0.91$\degr\, corresponding to 3.5 Mpc. The data set is reduced to 8.6 live hours due to a lack of appropriate off-source data. No significant signal is found. An upper limit is determined to be $F(>460\ \mathrm{GeV}) < 9.9\times 10^{-12} $ph. s$^{-1}$ cm$^{-2}$ (13.6\% Crab flux) for $\Gamma=2.1$ and $F(>460\ \mathrm{GeV}) < 1.1\times 10^{-11} $ph. s$^{-1}$ cm$^{-2}$ (15.1\% Crab flux) for $\Gamma=2.3$.

The next analysis probes the {\em 1 Mpc region}, corresponding to $\theta_4=0.26^{\circ}$. Again no signal is found and upper limits are derived: $F(> 1 \mathrm{TeV}) < 3.2 \times 10^{-13} $ph. s$^{-1}$ cm$^{-2}$ (1.4\% Crab flux) for $\Gamma=2.1$ and $F(> 1 \mathrm{TeV}) < 3.3 \times 10^{-13} $ph. s$^{-1}$ cm$^{-2}$ (1.4\% Crab flux) for $\Gamma=2.3$.

The 95\% X-ray containment (Perkins et al. \cite{perkins06}) region is $510$ kpc (based on parameters from Mohr et al. \cite{mohr99}), corresponding to an angular cut of $\theta_5 = 0.13^{\circ}$. No signal is found and the upper limit on integral flux is $F(> 460\ \mathrm{GeV}) < 3.4 \times 10^{-13} $ph. s$^{-1}$ cm$^{-2}$ (0.5\% Crab flux) for $\Gamma=2.1$ and $F(> 460\ \mathrm{GeV}) < 3.6 \times 10^{-13} $ph. s$^{-1}$ cm$^{-2}$ (0.5\% Crab flux) for $\Gamma=2.3$.

\begin{figure}[htbp] 
  \resizebox{\hsize}{!}{\includegraphics{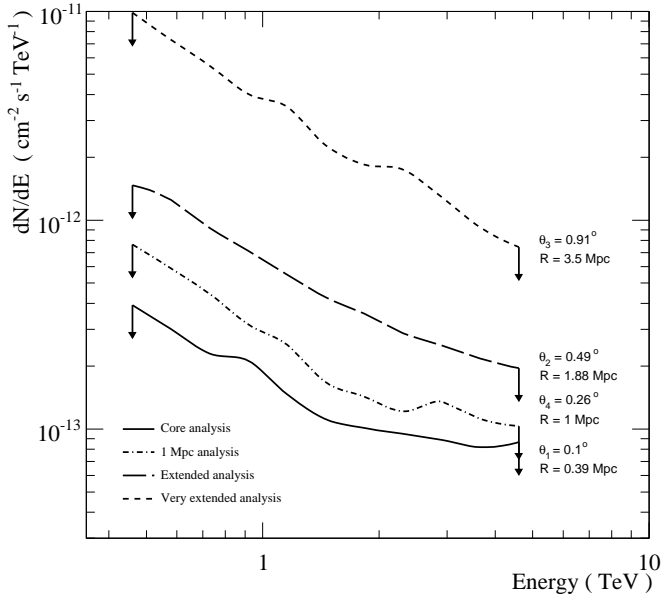}}
  \caption{The H.E.S.S. 99.9\% upper limits on the integral flux assuming different sizes of the emission region. The curves are obtained in the same way as in Fig. \ref{fig_ul}. All curves were produced assuming a power-law with a spectral index of $\Gamma=2.1$. The upper limit for $\theta_5 = 0.13^{\circ}$ is not shown for the sake of readability. It lies however very close to the curve of {\em point source analysis}. }\label{fig_a85_ul}
\end{figure}

The upper limits on the integral fluxes are plotted in Fig. \ref{fig_a85_ul}.
An independent calibration and analysis method combining
a semi-analytical shower model and the Hillas analysis
(\cite{deNaurois05}) was used as a cross-check for both clusters, yielding consistent
results.

\section{Discussion}\label{section_discussion}

In the following, the production mechanisms of CRs are discussed and applied to estimate the total energy of non-thermal particles in the galaxy clusters (Section \ref{section_expectations}). The upper limits in the TeV band obtained in the Section \ref{section_results} are used to experimentally constrain the non-thermal content of \abel\ and Abell~85 in Section \ref{section_constraints}.

\subsection{Estimates of energy in non-thermal particles}\label{section_expectations}

For the understanding of the non-thermal history of galaxy clusters, the important parameter is the fraction of energy that is contained in non-thermal particles with respect to the thermal energy of the ICM (E$_\mathrm{th}$). The total energy of non-thermal particles is dominated by the hadronic component, since no losses are expected for particles with energies of up to $\sim$10$^{15}$ eV during the Hubble time. In contrary to that VHE electrons suffer from energy losses and do not accumulate in galaxy clusters. Therefore, for all the following calculations only hadronic CRs are considered.    

In clusters of galaxies CRs are accelerated by two distinct mechanisms: externally via accretion shocks and hierarchical merger events, or via internal mechanisms such as supernova driven galactic winds and AGN outbursts. The energy in CRs produced by external processes will be proportional to the thermal energy of the ICM if it is assumed that the thermal plasma was shock-heated during the assembly of the cluster and that the CRs were accelerated in the same large scale shocks. The energetics of CRs resulting from supernova related processes will depend on the kinetic energy released by the supernova activity during the entire Hubble time, and thus on the total number of supernovae exploding in the cluster volume (V\"olk et al. \cite{voelk96}). Finally the contribution from AGNs will be constrained by the kinetic power of the outbursts and the timescale of activity. The total non-thermal energy is given by: 

\begin{equation}
\label{e1}
  E_\mathrm{nonth} = \epsilon_\mathrm{ext} E_\mathrm{kin}^\mathrm{ext} +  \epsilon_\mathrm{int} E_\mathrm{kin}^\mathrm{SN} + P_\mathrm{AGN} t_\mathrm{active}
\end{equation}

In this equation, $E_{\rm nonth}$ is the total energy in CRs, $\epsilon_{\rm ext}$ is the efficiency of the accretion shock in accelerating CR particles, $E_{\rm kin}^{\rm ext}$ is the kinetic energy of the accretion process which assembled the present day cluster, $\epsilon_{\rm int}$ is the efficiency of supernovae and galactic winds in accelerating particles, $ E_{\rm kin}^{\rm SN}$ is the combined kinetic energy provided by all supernovae exploding in the cluster volume, $P_{AGN}$ is the mean power of AGN activity and $t_{\rm active}$ is the combined duration of all AGN outbursts. Values for the particle acceleration efficiency of shocks in galaxy clusters are uncertain and expected typical numbers are used for these estimates. The contributions to $E_{\rm nonth}$ for the three described mechanisms are briefly estimated in the following. 

\textit{Accretion shocks:} In galaxy clusters the accretion process is the dominant mechanism that heats the ICM. Intrinsic processes such as supernovae provide only a minor contribution to the thermal energy of the ICM (Renzini \cite{renzini03}). Hence the thermal energy is a good measure of the kinetic energy $E_{\rm kin}^{\rm ext}$ of the infall of the cluster building blocks.
The thermal energy of the ICM can be determined by measuring the temperature and the density profile of the ICM. Here for Abell~496, a uniform temperature of 4.7 keV throughout the cluster volume and a $\beta$ density profile for the ICM is adopted (numbers from Markevitch et al. \cite{markevitch99}). A value of $4.3 \times 10^{62}$ erg for the thermal energy is computed. For Abell~85, a temperature of 7 keV and the density profile of Pfrommer \& En\ss lin (\cite{pfrommer04}) is used. With these input values the thermal energy of the ICM in Abell~85 is found to be $2.4 \times 10^{63}$ erg. The value of $\epsilon_{\rm ext}$ should be close to 0.1 (e.g. Dorfi \cite{dorfi91}) for strong shocks, but authors using nonlinear simulations claim that it could be as large as 0.5 (Kang et al. \cite{kang02}). However, if the accreted material is already hot, the sound velocity in such a medium is large. Hence typical Mach numbers of shocks in such media are small and these weak shocks are less efficient in accelerating particles (e.g. Dorfi \& V\"olk \cite{dorfi96}). This fact could reduce the value of $\epsilon_{\rm ext}$ in case of accretion of hot gas during sub-cluster mergers significantly. Keshet et al. (\cite{keshet04}) have argued that only a fraction of 8 -- 17\% of the baryonic material of a galaxy cluster is accreted in strong shocks. This accretion is also the contribution to $E_{\rm th}$ that is accompanied by effective CR acceleration because the rest of the material is assembled in weak shocks with inefficient non-thermal particle production. Here it is assumed that 10\% of the ICM is accreted in strong shocks with $\epsilon_{\rm ext} = 0.3$. Consequently, in this scenario the energy in the non-thermal component will be 3\% of the thermal energy for both clusters. 

\textit{Supernova activity:} To constrain the energetics of internal processes in the cluster, the total cluster iron mass can be used to  estimate the total number of supernovae (and thus the combined kinetic energy) that occurred in the cluster volume over a Hubble time. Supernovae come in general in two distinct classes which both feature about the same mean kinetic energy but differ considerably in the amount of iron produced: type Ia supernovae generate on average 0.7 M$_{\odot}$ of iron whereas the iron yield of core collapse supernovae is 0.07 M$_{\odot}$ (see Renzini \cite{renzini03} for a review). The ejecta of supernovae Ia can be distinguished from the ejecta of core collapse supernovae by the ratio of iron to oxygen (Renzini \cite{renzini93}). By knowing the relative contribution of supernovae Ia and core collapse supernovae, respectively, and using the mean iron mass produced by each class of events, it is possible to evaluate the total number of supernovae that have distributed the observed iron. From the Fe/O ratio found in Abell~496 and Abell~85 it is evident that type Ia supernovae contribute significantly to the chemical enrichment of the central cooling region (see Tamura et al. \cite{tamura04}) but are unimportant for the total iron mass in the entire cluster (De Grandi et al. \cite{degrandi04}). Hence it is assumed that the observed iron mass of $2.1 \times 10^{10}$ M$_{\odot}$ (Abell~496) and  $5.0 \times 10^{10}$ M$_{\odot}$ (Abell~85) (De Grandi et al. \cite{degrandi04}) was entirely generated by core collapse supernovae, and therefore an iron production mass of 0.07 M$_{\odot}$ per supernovae (Renzini \cite{renzini93}) is adopted. As a result of these considerations it is found that $3 \times 10^{11}$ supernovae (Abell~496) and $7.1 \times 10^{11}$ supernovae (Abell~85) are necessary to enrich the ICM to the observed level with iron. By adopting the canonical energy per supernova to be $10^{51}$ erg, a total energy of internal processes of $3\times 10^{62}$ erg (Abell~496) and $7.1\times 10^{62}$ erg (Abell~85) is derived. This energy is injected into the ICM in the form of kinetic energy, thermal energy and CRs, and furthermore, depending on the environment of the supernova explosions, parts of it will also go to radiation losses (Dorfi \cite{dorfi91}, Thornton et al. \cite{thornton98}). When assuming that 10\% of the initial kinetic energy of the supernovae is converted into CRs ($\epsilon_{\rm int}$ = 0.1 e.g. Dorfi \cite{dorfi91}) either in the supernova shocks themselves or in the termination shocks of supernova-driven galactic winds, it is found that supernova activity produces a component of high energy particles with about 7\% (Abell~496) and 3\% (Abell~85) of the thermal energy of the ICM.
Note that $\epsilon_{\rm int}$ can be smaller than 0.1 if a large number of supernovae explode directly in the hot ICM (Domainko et al. \cite{domainko04}, Zaritsky et al. \cite{zaritsky04}) because in this case SN shocks are less efficient in accelerating CRs (Dorfi \& V\"olk \cite{dorfi96}).

\textit{AGN activity:} AGNs are generally considered to be a major source of CRs in galaxy clusters (e.g. En\ss lin et al. \cite{ensslin97}, Aharonian \cite{aharonian02}, Hinton et al. \cite{hinton07}). In cooling core clusters, the AGN activity is usually dominated by a powerful central AGN. In the case of Abell~496 no AGN -- ICM interaction is observed (Dunn \& Fabian \cite{dunn06}) but this could simply be due to a low activity period of the central galaxy at present. 
It has to be noted that in several galaxy clusters, buoyantly rising bubbles filled with radio emitting relativistic electrons are found as remnants of past AGN outbursts (e.g. McNamara et al. \cite{mcnamara01}, Fabian et al. \cite{fabian02}), which is also not seen in the cluster Abell~496 (Dunn \& Fabian \cite{dunn06}).
The situation is different for the cluster Abell~85. In this cluster bubbles filled with non-thermal electrons injected by a past AGN outburst can be seen in the ICM. These bubbles are about 10$^7$ years old and the energy that is necessary to inflate these bubbles is $\sim 10^{58}$ erg (Dunn et al. \cite{dunn05}).
In general, AGN can act in repetitive outbursts with up to 100 cycles over the lifetime of the galaxy cluster. In each cycle the AGN is active for $\sim$10\% of the time (McNamara et al. \cite{mcnamara01}). One may estimate the contribution of potential AGN activity in the past to the non-thermal particle component in this cluster. If it is assumed that a powerful AGN injects 10$^{45}$~erg/s of hadronic CRs in the ICM and is active for 10\% of the cluster lifetime (which is essentially the Hubble time) then this mechanism will distribute about $3 \times 10^{61}$~ergs of CRs. Hence in an optimistic scenario, AGNs have the ability to provide non-thermal particles with an energy of  7\% (Abell~496) and 1\% (Abell~85) of the thermal energy of the ICM to the galaxy cluster.

In summary, it is found that all three processes (accretion shocks, supernova activity and AGN outbursts) may contribute a comparable share to $E_{\rm nonth}$ and it is further found that in total about 17\% (Abell~496) and 7\% (Abell~85) of the thermal energy of the cluster can be in the form of CRs. As stated above these numbers have large uncertainties and typical values are given.  Additionally it is shown that the contribution to $E_{\rm nonth}$ by internal processes is more important for less massive, cooler clusters.

\subsection{Probing the non-thermal energy content of the cluster}\label{section_constraints}

 
From the upper limits on the integral VHE gamma-ray flux obtained by the presented H.E.S.S. observations, it is possible to estimate the upper limit on the total energy of all hadronic CRs accelerated in these clusters within their lifetime for energies that are large enough for $\pi^0$ production, but where particles are still confined in the system.

The following estimates of the upper limits on the total energy in hadronic CRs are given for a radius of 1 Mpc (0.4$^{\circ}$ for Abell~496 and 0.26$^{\circ}$ for Abell~85) for both clusters (see Table \ref{nonth}). For Abell~496, according to Markevitch et al. (\cite{markevitch99}), the temperature of the ICM is 4.7 keV and the total gas mass of the thermal ICM is $3.2\times 10^{13}$ M$_{\odot}$ within a radius of 1 Mpc. The typical uncertainty of the quantities obtained from X-ray observations is about 10\%.
From the H.E.S.S. observations the upper limits presented in Sec. \ref{section_results} are found in this volume. For Abell~85 a temperature of 7 keV (Durret et al. \cite{durret05}) and a gas mass of $1.2\times 10^{14}$ M$_{\odot}$ (following the density profile of Pfrommer \& En\ss lin \cite{pfrommer04}) is used. From the H.E.S.S. observations the upper limits presented in Sec. \ref{section_results} are found within a radius of 1 Mpc for Abell~85.

Since the distribution of CRs in galaxy clusters is not known, the implications of these upper limits on different scenarios of CR injection and propagation are investigated. First, a model with a constant CR  density throughout the entire cluster is applied. This distribution is supported by a scenario where CRs are accelerated in accretion shocks at the outskirts of the cluster and are mainly transported to the higher gas density regions of the cluster center by bulk motions of the ICM caused by hierarchical merger events (see e.g. Miniati \cite{miniati03} for a discussion on the possible distributions of CRs in galaxy clusters).

As a second scenario, it is assumed that the CR density follows the large scale density distribution of the thermal gas excluding the central cooling region. This is motivated by a scenario where the CRs are mainly injected by cluster galaxies which are more concentrated towards the cluster center. 

Finally, a somewhat more extreme model is adopted where the density in CRs in the central cooling region includes an additional spike, similar to the gas density.
It has to be noted that such a centrally peaked distribution of CRs should not form in clusters. Magneto-hydrodynamic instabilities are expected to lead to a de-mixing of gas and relativistic particles where relativistic particles with a softer equation of state will rise buoyantly to larger cluster radii and will show a less centrally concentrated distribution than the thermal gas (e.g. Parker \cite{parker66}, Breitschwerdt et al. \cite{breitschwerdt93}, Chandran \& Dennis \cite{chandran06}). Furthermore, since the central cooling region is much smaller than the whole galaxy cluster, CRs can leave this region also due to diffusion.

In addition to the numbers computed for a radius of 1 Mpc, upper limits of $E_{\rm nonth}/E_{\rm th}$ for the whole cluster are calculated with the upper limits on the gamma radiation found for a radius of the clusters that contains 95\% of the expected gamma-ray emission (Perkins et al. \cite{perkins06}). The 95\% containment radius is computed with the assumption that the density in CRs follows the gas density including the high density cooling region (see Perkins et al. \cite{perkins06}). The radius of 95\% containment determined in this way is 1 Mpc for Abell~496 and 510 kpc for Abell~85. Values of $E_{\rm nonth}/E_{\rm th}$ obtained with this specific method for both clusters are again given in Tab. \ref{nonth}. It has to be noted that limits on $E_{\rm nonth}/E_{\rm th}$ derived in this way are particularly small for clusters with a highly peaked gas density. This results from deep upper limits on the gamma-ray brightness of the cluster due to a small spatial integration region, which can be adopted for the gamma-ray analysis for this case. However it has to be further noted that the value of $E_{\rm nonth}/E_{\rm th}$ obtained in this way only holds for the assumption that the density of CRs is as peaked towards the cluster center as the gas density. This scenario seems to be disfavored by theoretical considerations as has been discussed in the previous paragraph.

For the calculations of the upper limits of the non-thermal component of the cluster, the distribution of the thermal gas excluding the central cooling region (3$'$ = 118 kpc) found by Markevitch et al. (\cite{markevitch99}) ($\beta$ model) and the density profile obtained by Durret et al. (\cite{durret00}) for the cluster Abell~496 including the cooling core (CC included) is used. In the case of Abell~85 the distribution of CRs is assumed to follow the single large scale $\beta$ profile from  Pfrommer \& En\ss lin (\cite{pfrommer04}) ($\beta$ model) and to trace the combined double $\beta$ profile with an additional component in the central high density cooling region (CC included) (again from  Pfrommer \& En\ss lin \cite{pfrommer04}). For computing the gamma-ray luminosities, the production rate of gamma rays per unit volume is integrated over the entire galaxy cluster assuming spherical symmetry. Calculations are obtained for the hadronic channel adopting the gamma-ray emissivity according to Drury et al. (\cite{drury94}). Results are given in Table \ref{nonth} for a spectral index of the CR protons of 2.1 and 2.3. A hard spectrum is expected in galaxy clusters since no losses of CRs should occur there, and therefore the observed spectrum should have the same energy distribution as the primarily accelerated CRs. Depending on the model, limits on $E_{\rm nonth}/E_{\rm th}$ range from 0.40 to 1.12 for Abell~496 and from 0.03 to 0.15 for Abell~85 when a spectral index of 2.1 is assumed.

\begin{table}[h]
\caption{Upper limits on the ratio of energy in the non-thermal component with respect to the thermal energy of the ICM ($E_{\rm nonth}/E_{\rm th}$) for different spatial distributions of the CRs in the cluster. All numbers are given for a radius of 1 Mpc with the exception of the numbers for the 95\% containment radius (see main text). Particles of the non-thermal component have much larger energies ($>$ 1 GeV) than can be obtained thermally and are responsible for the gamma ray production whereas the thermal component can be observed in X-rays. A value of $E_{\rm nonth}/E_{\rm th} > 1$ means that the energy of the non-thermal component exceeds the energy of the thermal component. This configuration is unrealistic since it would require that shocks are more efficient in accelerating particles than in heating up the shocked medium. Therefore, the presented observations can not constrain any models for such a case. The uncertainties on the values given in this table is about 40\% due to the errors in the X-ray and gamma-ray observations.}

\begin{tabular}{l|l|cc}
\hline
\hline
 & & Abell~496 & Abell~85 \\
\hline
spectral & spatial & $E_{\rm nonth}/E_{\rm th}$ &  \\
index $\Gamma_{\gamma}$ & distribution of CRs & &  \\
\hline
2.1 & constant & 1.12 & 0.15 \\
 &  $\beta$ model & 0.51 & 0.08 \\
 & CC included & 0.40 & 0.06 \\
\hline
2.3 & constant & 5.66 & 0.75 \\
 & $\beta$ model & 2.56 & 0.40 \\
 & CC included & 2.03 & 0.30 \\
\hline
\hline
2.1 & CC included & & \\
 & 95\% containment & & \\
 & radius & 0.40 & 0.03 \\
\hline

\end{tabular}
\label{nonth}
\end{table}

Note that some models of other authors that concentrate on external production mechanisms predict a ratio of CR energy to gas thermal energy of up to $\sim$ 50\% (Miniati et al. \cite{miniati01}, Ryu et al. \cite{ryu03}). Hence the upper limits constrain models which favor a similarly large ratio of non-thermal to thermal energy. Limits obtained for Abell~85 are especially interesting. These limits are for a hard spectrum well within the prediction of the simple model developed in Sec. \ref{section_constraints} and exclude an unduly large component of hadronic CRs ($E_{\rm nonth}/E_{\rm th} > 0.15$). The non-detection of Abell~85 may even confirm the aforementioned theoretical arguments for which a very centrally concentrated distribution of CRs is disfavored. Indeed in this case, an upper limit of $E_{\rm nonth}/E_{\rm th} < 0.03$ can be derived which is smaller than model estimates (see Sec. \ref{section_expectations}). However it is important to mention that for a steeper spectrum of the CR protons, the limit on the energetics of the non-thermal component would be larger than these optimistic model predictions. 
For stronger constraints on the component of non-thermal particles in galaxy clusters, longer exposures are required.
In this context it should be noted that the sensitivity of H.E.S.S. scales with the square of the observation time and therefore, in order to reach a twice as sensitive limit on the energy in CRs as presented here, it is necessary to perform a four times longer observation.

Very recently, also constraining limits of E$_\mathrm{nonth}$/E$_\mathrm{th}$ at the few percent level for the galaxy cluster Abell~521 have been derived from radio observations (Brunetti et al. \cite{brunetti08}). With this approach the synchrotron emission of the secondary electrons produced by the proton - proton interactions is tested. The limits obtained in this way depend on the cluster magnetic field and are thus complementary to the limits derived from gamma-ray observations, which are independent of the magnetic field in the cluster.
 

\section{Conclusions}   
No significant point-like or extended gamma-ray flux F($>570$ GeV (Abell~496) and F($>460$ GeV) (Abell~85) has been found in the H.E.S.S. observations. Upper limits on the VHE gamma-ray flux from both clusters are derived.
With the presented results, it is possible to constrain the energy fraction in hadronic CRs. For the case where the energy density of the CRs follows the density of the thermal gas and where the CRs have a spectral index of 2.1 this is not more than 51\% (Abell~496) and 8\% (Abell~85) of the thermal energy. Especially for the case of Abell~85 the upper limits obtained with H.E.S.S. can already constrain model predictions on $E_{\rm nonth}/E_{\rm th}$. These limits are the best determined so far without a dependency on the magnetic field. However, for a uniform energy density of CRs or a softer spectrum of the particles, the limit on the fraction of energy in hadronic CRs versus thermal energy of the ICM is not constraining. Our theoretical considerations also suggest that the most promising clusters for VHE gamma-ray observations are those with the largest mass (M$_{\rm tot}$ $>$ 10$^{15}$ M$_{\odot}$) and highest temperatures (T $>$ 7 keV) and, as far as supernova activity is concerned, those with the largest iron mass (M$_{\rm Fe}$ $>$ 5$\times$10$^{10}$ M$_{\odot}$). Observations of this kind of galaxy cluster at a distance of $\sim$100 Mpc with the present generation of IACTs to a flux of $F(>1\ \mathrm{TeV}) < 10^{-12}$ ph. s$^{-1}$ cm$^{-2}$ within a radius of 1$^{\circ}$ will provide the ability to test models of the non-thermal hadronic component in galaxy clusters, which predict a considerable fraction ($\sim$10\%) of energy in CRs.

\begin{acknowledgements}
The support of the Namibian authorities and of the University of Namibia
in facilitating the construction and operation of H.E.S.S. is gratefully
acknowledged, as is the support by the German Ministry for Education and
Research (BMBF), the Max Planck Society, the French Ministry for Research,
the CNRS-IN2P3 and the Astroparticle Interdisciplinary Programme of the
CNRS, the U.K. Science and Technology Facilities Council (STFC),
the IPNP of the Charles University, the Polish Ministry of Science and 
Higher Education, the South African Department of
Science and Technology and National Research Foundation, and by the
University of Namibia. We appreciate the excellent work of the technical
support staff in Berlin, Durham, Hamburg, Heidelberg, Palaiseau, Paris,
Saclay, and in Namibia in the construction and operation of the
equipment.
\end{acknowledgements}


\end{document}